\documentclass[fleqn,10pt]{wlscirep}
\title{Chaos-guided Input Structuring for Improved Learning in Recurrent Neural Networks}

\author[1,*]{Priyadarshini Panda}
\author[1]{Kaushik Roy}
\affil[1]{School of Electrical \& Computer Engineering, Purdue University, West Lafayette, IN, USA 47907}

\affil[*]{Correspondence: pandap@purdue.edu}

\begin{abstract}
Anatomical studies demonstrate that brain reformats input information to generate reliable responses for performing computations. However, it remains unclear how neural circuits encode complex spatio-temporal patterns. We show that neural dynamics are strongly influenced by the phase alignment between the input and the spontaneous chaotic activity. Input structuring along the dominant chaotic projections causes the chaotic trajectories to become stable channels (or attractors), hence, improving the computational capability of a recurrent network. Using mean field analysis, we derive the impact of input structuring on the overall stability of attractors formed. Our results indicate that input alignment determines the extent of intrinsic noise suppression and hence, alters the attractor state stability, thereby controlling the network's inference ability.
\end{abstract}
\begin{document}
\flushbottom
\maketitle
%
%
\thispagestyle{empty}

\section*{Introduction}
Brain actively \textit{untangles} the input sensory data and fits them in behaviorally relevant dimensions that enables an organism to perform recognition effortlessly, in spite of variations \cite{dicarlo2012does, thorpe1996speed, dicarlo2007untangling}. For instance, in visual data, object translation, rotation, lighting changes and so forth cause complex nonlinear changes in the original input space. However, the brain still extracts high-level behaviorally relevant constructs from these varying input conditions and recognizes the objects accurately. What remains unknown is how brain accomplishes this untangling. 

Here, we introduce the concept of chaos-guided input structuring in a reservoir computing network that provides an avenue to untangle stimuli in the input space and improve the ability of a stimulus to entrain neural dynamics. Specifically, we show that the complex dynamics arising from the recurrent structure of a randomly connected reservoir \cite{rajan2006eigenvalue, kadmon2015transition, stern2014dynamics} can be used to extract an explicit phase relationship between the input stimulus and the spontaneous chaotic neuronal response. Then, aligning the input phase along the dominant projections determining the intrinsic chaotic activity, causes the random chaotic fluctuations or trajectories of the network to become locally stable channels or dynamic attractor states that, in turn, improve its' inference capability. In fact, using mean field analysis, we derive the effect of introducing varying phase association between the input and the network's spontaneous activity. Our results demonstrate that successful formation of stable attractors is strongly determined from the input alignment.
 
\section*{Model Description}
We describe the effect of chaos guided input structuring on a standard firing-rate based reservoir model of \textit{N} interconnected neurons. Specifically, each neuron in the network is described by an activation variable $x_i$ $\forall i= 1,2,...N$, satisfying
\begin{equation}
\tau dx_i/dt = -x_i + \sum_{j=1}^{N} W_{ij} r_j + W_{Input}I;\hspace{1mm}
z = \sum_{j=1}^{N} W_{Out} r_j
\end{equation}
where $r_i(t) = \phi(x_i(t))$ represents the firing rate of each neuron characterized by the nonlinear response function, $\phi(x) = tanh(x)$ and $\tau =10 ms$ is the neuron time constant. $W$ represents a sparse $N\times N$ recurrent weight matrix (with $W_{ij}$ equal to the strength of the synapse connecting unit $j$ to unit $i$) chosen randomly and independently from a Gaussian distribution with $0$ mean and variance, $g^2/p_c N$ \cite{van1996chaos, van1998chaotic}, where $g$ is the synaptic gain parameter and $p_c$ is the connection probability between units. 
The output unit $z$ reads out the activity of the network through the connectivity matrix, $W_{Out}$, with initial values drawn from a Gaussian distribution with 0 mean and variance $1/N$. The readout weights are trained using Recursive Least Square (RLS) algorithm \cite{laje2013robust, sussillo2009generating, jaeger2004harnessing}. The input weight matrix, $W_{Input}$, is drawn from a Gaussian distribution with zero mean and unit variance. The external input, $I$, is an oscillatory sinusoidal signal, $I = I_0 cos(2\pi ft + \chi)$, with amplitude $I_0$, frequency $f$, that is the same for each unit $i$. Here, we use a phase factor $\chi$ chosen randomly and independently from a uniform distribution between $0$ and $2\pi$. This ensures that the spatial pattern of input is not correlated with the recurrent connectivity initially. Through input alignment analysis, we then obtain the optimal phases to project the inputs in the preferred direction of the network's spontaneous activity. In all our simulations (without loss of generality), throughout the paper we have assumed, $p_c = 0.1, N=800, g=1.5, f= 10 Hz$. 

\begin{figure*}
\includegraphics[width=\linewidth]{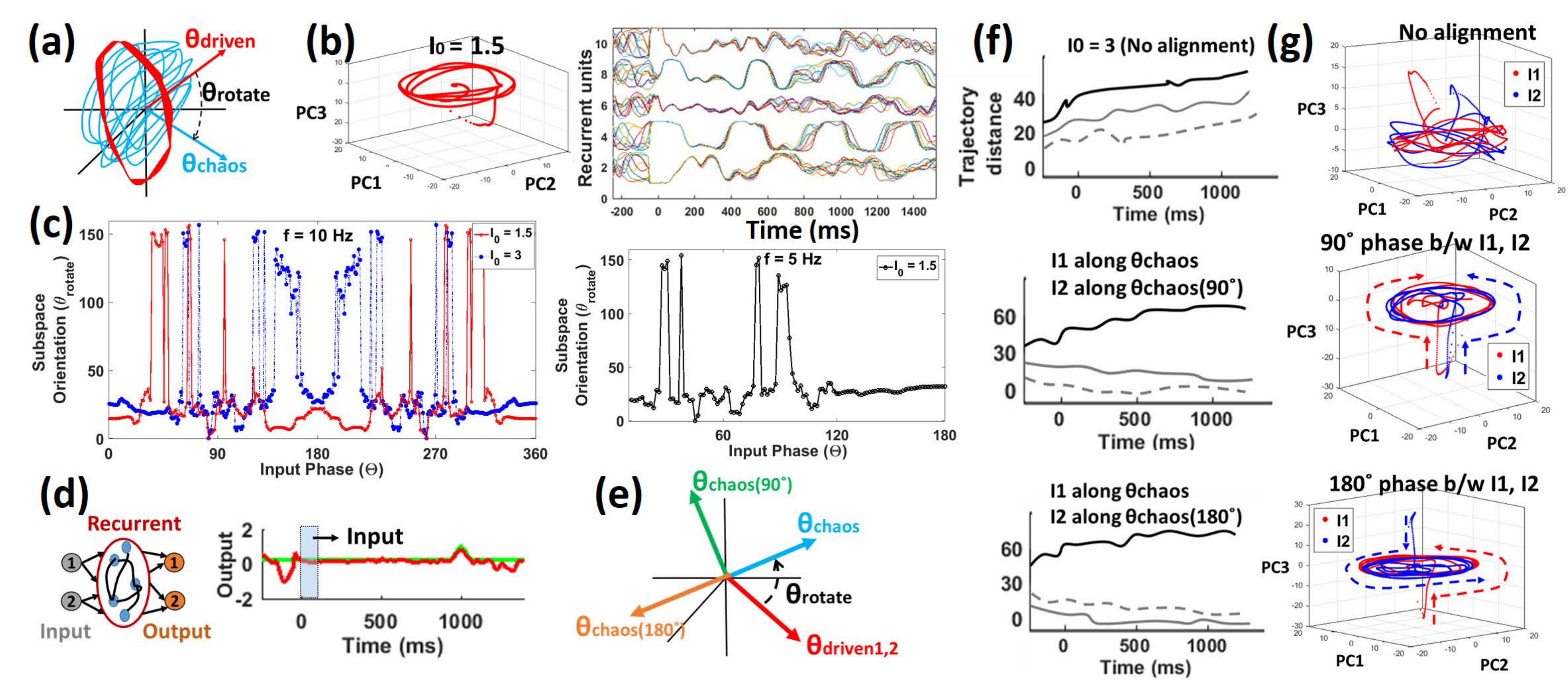}
\caption{\label{fig2} (a) Cartoon depicting the angle between subspace defined by the first two PCs of chaotic activity (blue) and input driven activity (red). (b) [Left] Projections of the reservoir activity in the 3D space of PC Vectors 1, 2, 3 - / [Right] Trajectories of 5 reservoir neurons across 10 different trials - driven by input rotated by $\theta_{rotate}$. (c) Relationship between input temporal phase ($\Theta$) and the orientation of driven activity with respect to chaotic subspace ($\theta_{rotate}$) for varying input amplitude and frequency. (d) A reservoir framework [Left] stimulated by a brief sinusoidal input ($t=0 - 50 ms$) trained to generate a timed output response shown in [Right]. (e) Cartoon showing the different angles in the chaotic subspace along which the inputs can be aligned. (f) Euclidean distance between trajectories of same (and different) inputs plotted for different orientation of $I_1, I_2$ in the chaotic subspace. (g) Projections of the reservoir activity in the 3D space of PC Vectors 1, 2, 3 corresponding to the two inputs $I_1, I_2$ for different alignment conditions.} 
\vspace{-4mm}
\end{figure*}

\section*{Subspace Alignment}
First, we ask the question how is the subspace of input driven activity aligned with respect to the subspace of spontaneous activity of a recurrent network. Using Principal Component Analysis, we observed that the input-driven trajectory converges to a uniform shape becoming more circular with increasing input amplitude \cite{S1ab}. We utilize the concept of principal angles, introduced in \cite{rajan2010inferring, ipsen1995angle}, to visualize the relationship between the chaotic and input driven (circular) subspace. Specifically, for two subspaces of dimension $D_1$ and $D_2$ defined by unit principal component vectors (that are mutually orthogonal) ${V_1}^a$, for $a = 1,2,...D_1$ and ${V_2}^b$, for $b = 1,2,...D_2$, the angle between them is 
\begin{equation}
\theta = arccos(min(SingularValueOf({V_1}^a . {V_2}^b)))
\label{eq3}
\end{equation}

Fig. \ref{fig2} (a) schematically represents the angle between the circular input driven network activity and the irregular spontaneous chaotic activity. Here, $\theta_{chaos}$ (and $\theta_{driven}$) refers to the subspace defined by the first two Principal Components (PCs) of the intrinsic chaotic activity (and periodic driven activity). It is evident that rotating the circular orbit by $\theta_{rotate}$ will align it along the chaotic trajectory projection. 
We observe that aligning the inputs in directions (along dominant PCs) that account for maximal variance in the chaotic spontaneous activity facilitates intrinsic noise suppression at relatively low input amplitudes, thereby, allowing the network to produce stable trajectories. For instance, instead of using random phase input, we set $I = I_0 cos(2\pi ft + \Theta)$ and visualize the network activity as shown in Fig. \ref{fig2} (b). Even at lower amplitude of $I_0 =1.5$, we observe a uniform circular orbit (in the PC subspace) for the network activity that is characteristic of reduction in intrinsic noise and input sensitization. In fact, even after the input is turned off after $t=50 ms$, the neural units yield stable and synchronized trajectories with minimal variation across different trials (Fig. \ref{fig2} (b, Right)) in comparison to the random phase input driven network (of higher amplitude) in \cite{S1ab}. This shows the effectiveness of subspace alignment for intrinsic noise suppression. In addition, working in low input-amplitude regimes offers an additional advantage of higher network dimensionality \cite{S1c}, that in turn improves the overall discriminative ability of the network. Note, previous work \cite{rajan2010inferring, rajan2010stimulus} have shown that spatial structure of the input does not have a keen influence on the spatial structure of the network response. Here, we bring in this association explicitly with subspace alignment. 

$\Theta$, in the above analysis, is the input phase that corresponds to a subspace rotation of driven activity toward spontaneous chaotic activity. We observe that the temporal phase of the input contributes to the neuronal activity in a recurrent network. Fig. \ref{fig2} (c) illustrates this correlation wherein the input phase determines the orientation of the input-driven circular orbit with respect to the dominant subspace of intrinsic chaotic activity. For a given input frequency ($f =10 Hz$), input phase, $\Theta = 83.2^{\circ}$, aligns the driven activity ($\theta_{driven}$) along the chaotic activity ($\theta_{chaos}$) resulting in $\theta_{rotate} =0^{\circ}$ for varying input amplitude ($I_0 = 1.5, 3$). 
An interesting observation here is that the frequency of the input modifies the orientation of the evoked response that yields different input phases at which $\theta_{chaos}$ and $\theta_{driven}$ are aligned (refer to Fig. \ref{fig2} (c, Right)). We also observe that the subspace alignment is extremely sensitive toward the input phase in certain regions with abrupt jumps and non-smooth correlation. This non-linear behavior is a consequence of the recurrent connectivity that overall shapes the complex interaction between the driving input and the intrinsic dynamics. While this correlation yields several important implications for network modeling experiments, we utilize this behavior for subspace alignment. Consequently, in all our experiments, for a given  $\theta_{rotate}$, we find a corresponding input phase $\Theta$ that approximately aligns the input in the preferred direction. 

\section*{Impact of Input Structuring on Discriminative Capability}
Next, we describe the implication of input alignment along the chaotic projections on the overall learning ability of the network. 
First, we trained a recurrent network with two output units to generate a timed response at $t= 1s$ as shown in Fig. \ref{fig2} (d, Right). Two distinct and brief sinusoidal inputs (of 50 ms duration and amplitude $I_0 = 1.5$) were used to stimulate the recurrent network. The network trajectories produced were then mapped to the output units using RLS training (to learn the weights $W_{Out}$). Here, the network (after readout training) is expected to produce timed output dynamics at readout unit 1 or 2 in response to input $I_1$ or $I_2$, respectively. The network is reliable if it generates consistent response at the readout units across repeated presentations of the inputs during testing, across different trials. This simple experiment utilizes the fact that neural dynamics in a recurrent network implicitly encode timing that is fundamental to the processing and generation of complex spatio-temporal patterns. Note, in such cases of multiple inputs, values of both inputs are zero, except for a timing window during which one input is briefly turned on in a given trial. 

Since both the inputs, in the above experiment, have same amplitude and frequency dynamics, the circular orbit describing the network activity in the input-driven state (for both inputs) is almost similar giving rise to one principal angle ($\theta_{{driven}_{1,2}}$ in Fig. \ref{fig2} (e)) for the input subspace. To discriminate between the output responses for the two inputs, it is apparent that the inputs have to be aligned in different directions. One obvious choice is to align each input along two different principal angles defining the chaotic spontaneous activity (i.e. $I_1$ along $\angle PC1.PC2$ and $I_2$ along $\angle PC3.PC4$). Note, $\angle PC1.PC2$ denotes the angle $\theta$ calculated using Eqn. \ref{eq3}. Another approach is to align $I_1$ along $\angle PC1.PC2 \equiv \theta_{chaos}$ and $I_2$ along $\angle PC1.PC2 + 90^\circ \equiv \theta_{chaos,90^\circ}$ as shown in Fig. \ref{fig2} (e). We analyze the latter in detail as it involves input phase rotation in one subspace that makes it easier for formal theoretical analysis. 

To characterize the discriminative performance of the network, we evaluated the Euclidean distances (measured as $\sqrt{1/N\sum_{i=1}^{N}(r_{i,1}(t)-r_{i,2}(t))^2}$ , where $r_1(t)$ ($r_2(t)$) is the firing rate activity of the network corresponding to $I_1$ ($I_2$)) between the inter-/intra-input trajectories in response to different inputs ($I_1, I_2$), and to a slightly varied version of the same input (for instance, $I_{1,2} = (I_0 +\epsilon)cos(2\pi ft + \Theta_{1,2})$ where $\epsilon$ is a random number between $[0, 0.5]$) and $\Theta_1 (\Theta_2)$ is the input phase that aligns $I_1$ ($I_2$) along $\theta_{chaos}$ ($\theta_{chaos, 90^\circ}$). The inter-/intra-input trajectory distances are plotted in Fig. \ref{fig2} (f) for both scenarios-with and without input alignment. It is desirable to have larger inter-trajectory distance and small intra-trajectory distance such that the network easily distinguishes between two inputs while being able to reproduce the required output response even when a particular input is slightly perturbed. We observe that aligning the inputs in direction parallel and perpendicular to the dominant projections (Fig. \ref{fig2} (f, Middle)) increases the inter-trajectory distance compared to the non-aligned case (Fig. \ref{fig2} (f, Top)) while decreasing the intra-input trajectory separation. This further ascertains the fact that subspace alignment reduces intrinsic fluctuations within a network thereby enhancing its prediction capability. Note, without input alignment, the intrinsic fluctuations cannot be overcome with low-amplitude inputs ($I_0=1.5$). Hence, for fair comparison and to obtain stable readout-trainable trajectory in the non-aligned case, we use a higher input amplitude of $I_0 =3$.

We hypothesize that intrinsic noise suppression occurs as input subspace alignment along dominant projections (that account for maximal variance such as $PC1, PC2$) causes chaotic trajectories along different directions (in this case, along $\theta_{chaos}$, $\theta_{chaos,90^\circ}$)  to become locally stable channels or \textit{attractor states}. These attractors behave as potential wells (or local minima from an optimization standpoint) toward which the network activity converges for different inputs. Thus, the successful formation of stable yet distinctive attractors for different inputs are strongly influenced by the orientation along which the inputs are aligned. As a consequence of our hypothesis, depending upon the orientation of the input with respect to the dominant chaotic activity ($\theta_{chaos}$ in Fig. \ref{fig2} (e)), the extent of noise suppression will vary, for a particular trajectory, that will eventually alter the stability of the attractor states. To test this, we rotated $I_2$ (from $\theta_{chaos,90^\circ}$) further by $90^\circ$ ($\theta_{chaos,180^\circ}$ in Fig. \ref{fig2} (e)) and monitored the intra-trajectory distance. Note, $I_1$ and $I_2$ are anti-phase correlated in the chaotic subspace. In Fig. \ref{fig2} (f, Middle) corresponding to $90^\circ$ phase difference between ($I_1, I_2$), $I_2$ corresponds to a more stable attractor than $I_1$ since the intra-distance for the former is lower. In contrast, in Fig. \ref{fig2} (f, Right) corresponding to $180^\circ$ phase difference, $I_1$ turns out be more stable than $I_2$. Note, the $90^\circ, 180^\circ$ phase difference between $I_1, I_2$ (mentioned above and in the remainder of the paper) refers to the phase difference between the inputs in the chaotic subspace after subspace alignment using $\Theta$. For our analysis, $\Theta_1 = 83.2^\circ, \Theta_2 = 111^\circ$ yields $\sim 90^\circ$ phase between $I_1, I_2$ in chaotic subspace, while $\Theta_1 = 83.2^\circ, \Theta_2 = 263.2^\circ$ yields $\sim 180^\circ$ phase. 

In addition to the trajectory distance, visualizing the network activity in the 3-D PC space (Fig. \ref{fig2} (g)), also, shows the influence of input orientation (and hence the phase correlation) toward formation of distinct attractor states. Since $I_1, I_2$ are aligned in the subspace defined by $\angle PC1.PC2$, the 2D projection of the circular orbit onto PC1 and PC2 in both input aligned scenarios ($90^\circ$, $180^\circ$ phase) are comparable. However, the third dimension, PC3, marks the difference between the two input projections. In fact, the progress of the network activity as time evolves (shown by dashed arrows in Fig. \ref{fig2} (g)) follows a completely different cycle for the input aligned scenarios. The change in the overall rotation cycle from anti-clockwise ($I_2$ with $90^\circ$ phase, Fig. \ref{fig2} (g, Middle)) to clockwise ($I_2$ with $180^\circ$ phase, Fig. \ref{fig2} (g, Bottom)) can be viewed as an indication toward the altering of the attractor state stability. On the other hand, the non-aligned case with $I_0 =3$ yields incoherent and more random trajectory (Fig. \ref{fig2} (g, Top) representative of intrinsic noise. In order to get more coherent activity and to suppress the noise further, we need to increase the input amplitude to $I_0 \ge 5$ as shown in \cite{S1ab}.

\section*{Mean Field Analysis}
To explain the above results analytically, we use mean-field methods developed to evaluate the properties of random network models in the limit $N \rightarrow \infty$ \cite{rajan2010stimulus, sompolinsky1988chaos}. A key quantity in Mean Field Theory (MFT) is the average autocorrelation function that characterizes the interaction within the network as
\begin{equation}
C(\tau) = 1/N \sum_{i=1}^{N} <\phi (x_i(t)) \phi (x_i(t+\tau))>
\label{eq4}
\end{equation}
where $<>$ denotes the time average. The main idea of MFT is to replace the network interaction term in Eqn. 1 by Gaussian noise $\eta$ such that $\frac{d{x_i}^1}{dt} = -{x_i}^1 +\eta$, where $x_i = {x_i}^0 + {x_i}^1$ and ${x_i}^0 (t) = A cos(2\pi ft + \zeta)$ with $A = I_0/\sqrt{1+(2\pi ft)^2}$. Here, $\zeta$ incorporates the averaged temporal phase relationship between the reservoir neurons and the input induced by input subspace alignment, $\zeta(\theta_{rotate}) = \Theta$. The temporal correlation of $\eta$ is calculated self-consistently from $C(\tau)$.  For self-consistence, the first and second moment of $\eta$ must match the moments of the network interaction term. Thus, we get $<\eta_i (t)> = 0$ as mean of the recurrent synaptic matrix $<W_{ij}> = 0$. For calculating the second moment , we use the identity $<W_{ij}W_{kl}> = g^2 \delta_{ij} \delta_{kl}/N$ and obtain $<\eta_i (t) \eta_j(t+\tau)> = g^2C(\tau)$. Combining this result with the MFT noise-interaction based network equation yields
\begin{equation}
\frac{d^2 \Delta (\tau)}{d \tau^2} = \Delta(\tau) - g^2C(\tau)
\label{eq5}
\end{equation}
where $\Delta(\tau) = < {x_i}^1 (t) {x_i}^1 (t+\tau)>$. Eqn. \ref{eq5} resembles the Newtonian motion equation of a classical particle moving under the influence of force given by the right hand side of the equation. This force depends on $C$ that, in turn, depends on the input subspace alignment ($\zeta$) which directs the initial position of the particle (or state of the network $\Delta(0)$). From this analogy, it is evident that analyzing the overall potential energy function of the particle (or network) will be equivalent to visualizing the different attractor states formed in a network in response to a particular input stimulus. Thus, we formulated an expression for the correlation function (with certain constraints) using Taylor series expansion, that allows us to derive the force and hence the dynamics of the network under various input alignment conditions. 

The non-linear firing rate function $r(x) = \phi(x) = tanh(gx)$ can be expanded with Taylor series for small values of $g$, i.e. $g = 1 + \delta$, where $\delta$ denotes a small increment in $g$ beyond $1$. Note, $g=1+\delta$ satisfies the criterion, $g >1$ \cite{rajan2006eigenvalue, sompolinsky1988chaos}, to operate the networks in chaotic regime. Also, the overall network statistics does not change with $g$ being expressed as a gain factor in the firing-rate function instead of overall synaptic strength. Using $tanh(gx) \simeq gx-1/3g^3x^3 +2/15g^5x^5$, we can express $C(\tau)$ from Eqn. \ref{eq4} as  $C = 1/2g^2A^2cos(\omega \tau +2\zeta) + l(g^2-2g^4m) + 2/3g^6l^3$, where $m=\Delta(0), l=\Delta(\tau), \omega = 2\pi f$. Now, we can express Eqn. \ref{eq5} as $\frac{d^2l}{d \tau^2} = l- C$. Writing $l=k\delta$ due to the small limit of $g$, Eqn. \ref{eq5} simplifies to
\begin{equation}
\frac{d^2k}{dt^2} = -G +  n k -2/3k^3
\label{eq10}
\end{equation}

where $G = g^2A^2cos(\omega \tau +2\zeta)/(2\delta^3) $ and $n$ is a parameter defined in terms of $m, \delta$. Note, supplementary section \cite{S2} provides a detailed derivation of Eqn. \ref{eq10} and comments about the assumptions on initial conditions. Note, Eqn. \ref{eq10} is an approximate version of Eqn. \ref{eq5} that depicts network activity in the manner of Newtonian motion independent of all intrinsic time (or averaging parameters) while taking into account the influence of input alignment. Now, we can express the potential of the network driven by a force, $F$, equivalent to the right hand side of Eqn. \ref{eq10} as
\begin{equation}
V= -\int{Fdk} = Gk-n/2k^2+k^4/6
\label{eq11}
\end{equation}
We solve Eqn. \ref{eq10}, \ref{eq11} with initial conditions $k(0) =1$, $\dot{k}(0) =0$ and monitor the change in force, $F$, and potential, $V$, for different values of $G$. First, let us examine the attractor state formation when there is no input stimulus (i.e. $G=0$) by visualizing the potential $V$. For $G=0$, the expressions for force and potential become

\begin{equation}
F(k) = nk-2/3k^3 ;\hspace{2mm} V(k) = -nk^2/2+k^4/6
\label{eq12}
\end{equation}

\begin{figure}
\centering
\includegraphics[width=0.6\linewidth]{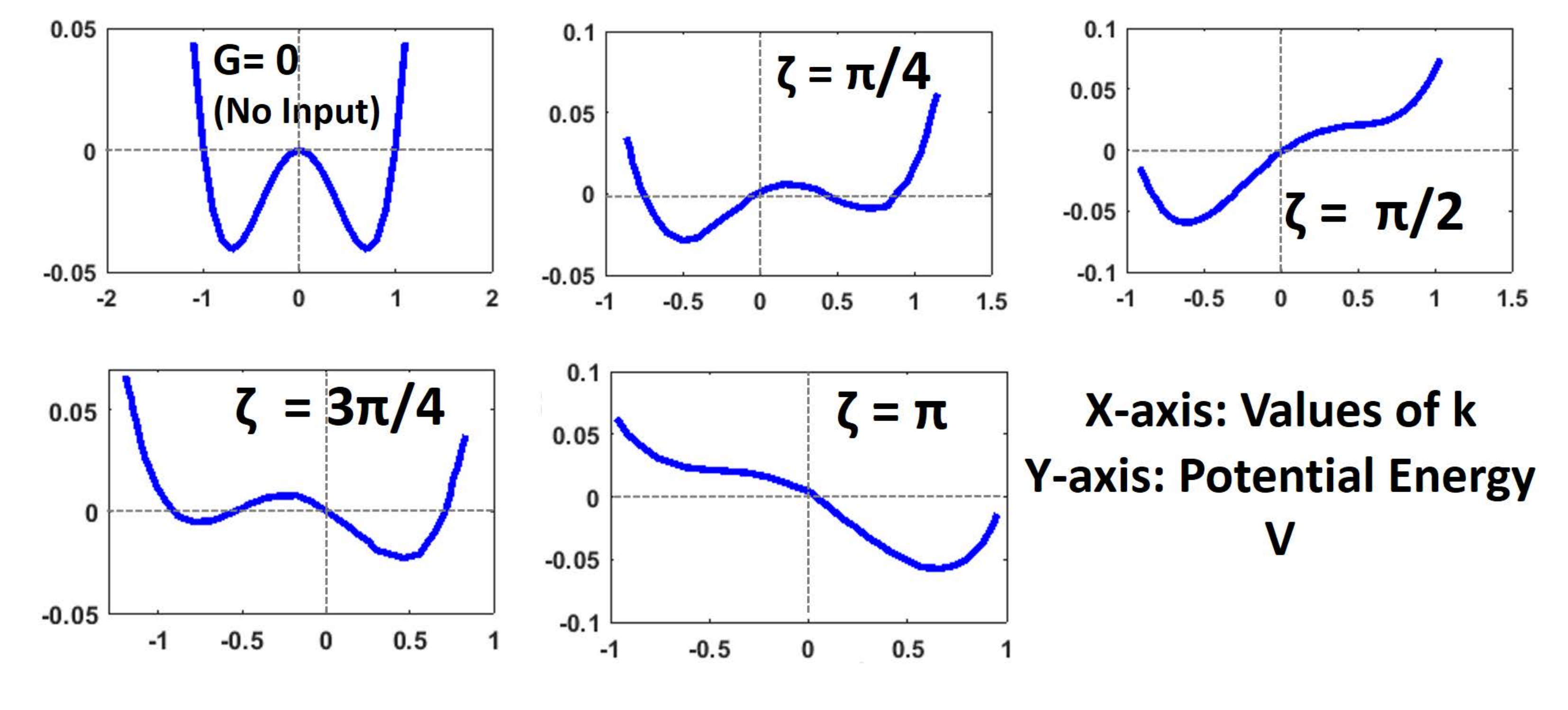}
\vspace{-4mm}
\caption{\label{fig4} Evolution of potential energy (and hence attractor state formation) by varying the input $G$ as a function of $\zeta$}
\vspace{-4mm}
\end{figure}

Fig. \ref{fig4} shows the evolution of potential energy as $k$ varies for different $G$. Since external input $G=0$, the network dynamics is chaotic that results in the formation of potential wells that are both equally stable. The network activity will thus converge to any one of these wells (that can be interpreted as attractor states) depending upon the initial state or starting point. This supports the observation in \cite{S1ab} that a network with no input yields chaotic activity with incoherent and irregular trajectory for every trial. For nonzero G, the force (and potential) equation will be dependent on $\zeta$ since $G \simeq cos(\omega \tau +2\zeta)$. For different values of $\zeta$, we solved for $V$ (Eqn. \ref{eq11}) numerically and plotted the potential evolution as shown in Fig. \ref{fig4}. For $\zeta =\pi/4$, the potential well is more attractive on the left end. This validates the fact that intrinsic fluctuations are suppressed in the presence of an input. For $\zeta =\pi /2$, the left attractor becomes more stable. Changing $\zeta$ further shows that the potential well on the right end becomes more stable. This result confirms that input subspace alignment with respect to the initial chaotic state influences the overall stability and convergence capability of a recurrent network. 

The fact that stability corresponding to different attractor states ($\zeta =\pi /2, \pi$) arises, qualifies our earlier hypothesis that input orientation with respect to the chaotic subspace alters the attractor state stability, corroborating the result of Fig. \ref{fig2} (f). Finally, we illustrate the effectiveness of input alignment on a complex motor pattern generation task with reliable generation of learnt handwritten patterns over multiple trials, even in presence of perturbations \cite{laje2013robust}. The detailed analysis and results are shown in the supplementary \cite{S3}. 

Note, we solved Eqn. \ref{eq11} by setting some initial and boundary value conditions on $k$ and by iterating over different $n$ until we reached a steady state solution. Changing these conditions will result in a completely new set of $\zeta$ values (different from those in Fig. \ref{fig4}). Nevertheless, we will observe a similar evolution of the potential well and change in attractor state stability as Fig. \ref{fig4}. Furthermore, the MFT calculations use $\zeta$ to denote a functional relationship between subspace alignment and input phase that eventually affects the attractor state stability. In the future, we will examine the real-time evaluation of $\zeta$ and its' impact on the analytical studies. Finally, the constraint under which we derive the potential energy functions and show the altering of attractor state is $g = 1 + \delta$. We expect all our results to be valid for large $g$ as well since Eqn. \ref{eq5} (that was simplified with Taylor expansion) still remains unchanged. 

\section*{Conclusion}
Models of cortical networks often use diverse plasticity mechanisms for effective tuning of recurrent connections to suppress the intrinsic chaos (or fluctuations) \cite{laje2013robust,panda2017learning}. We show that input alignment alone produces stable and repeatable trajectories, even, in presence of variable internal neuronal dynamics for dynamical computations. Combining input alignment with recurrent synaptic plasticity mechanism can further enable learning of stable correlated network activity at the output (or readout layer) that is resistant to external perturbation to a large extent. Furthermore, since input subspace alignment allows us to operate networks at low amplitude while maintaining a stable network activity, it provides an additional advantage of higher dimensionality. A network of higher dimensionality offers larger number of disassociated principal chaotic projections along which different inputs can be aligned \cite{S1c}. Thus, for a classification task, wherein the network has to discriminate between 10 different inputs (of varying frequencies and underlying statistics), our notion of \textit{untangling} with chaos-guided input structuring can, thus, serve as a foundation for building robust recurrent networks with improved inference ability. Further investigation is required to examine which orientations specifically improve the discrimination capability of the network and the impact of a given alignment on the stability of the readout dynamics around an output target. In summary, the analyses we present suggest that input alignment in the chaotic subspace has a large impact on the network dynamics and eventually determines the stability of an attractor state. In fact, we can control the network's convergence toward different stable attractor channels during its voyage in the neural state space by regulating the input orientation. This indicates that, besides synaptic strength variance \cite{rajan2006eigenvalue}, a critical quantity that might be modified by modulatory and plasticity mechanisms controlling neural circuit dynamics is the input stimulus alignment. 



\subsection*{Acknowledgments}
P.P. and K.R. are supported in part by Center for Brain-inspired Computing (C-BRIC), an SRC and DoD sponsored center, the Semiconductor Research Corporation, the NSF, Intel Corporation and by the DoD Vannevar Bush Fellowship.

\section*{Supplementary Material}
\section{PCA for examining Network Activity}
\begin{figure}
\includegraphics[width = \linewidth]{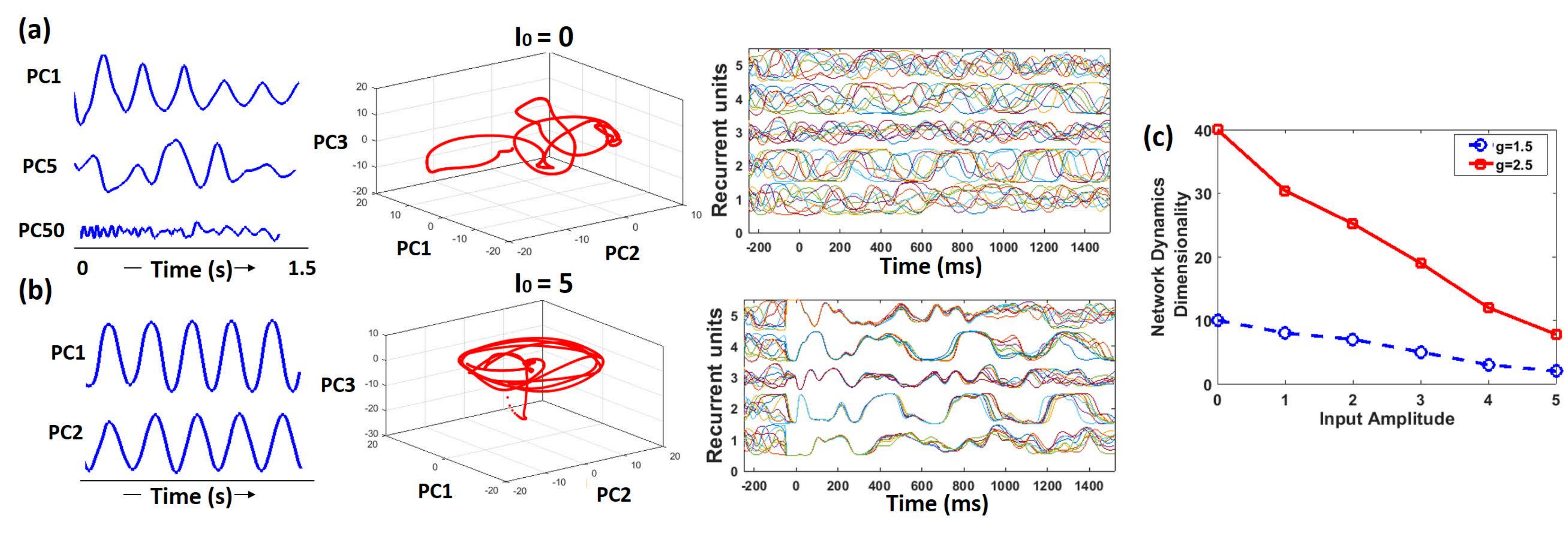}
\caption*{\textbf{Figure S1} (a) [Left] For zero input, projections of the chaotic spontaneous activity onto PC vectors 1, 5, 50 , [Middle] Visualization of the chaotic trajectory in 3D subspace composed of dominant PC vectors 1,2,3 that account for significant variance in network activity, [Right] Trajectories of 5 reservoir neurons across 10 different trials. (b) Same as panel (a), but for non-chaotic input driven activity with input amplitude $I_0 =5$. (c) Effective dimensionality of the network at different input amplitudes for g=1.5, g=2.5.}
\end{figure}
To examine the structure of the recurrent network’s representations, we visualize and compare the neural trajectories in response to varying inputs using Principal Component Analysis (PCA) \cite{rajan2010inferring}. The network state at any given time instant can be described by a point in the \textit{N}-dimensional space with coordinates corresponding to the firing rates of the \textit{N} neuronal units. With time, the network activity traverses a trajectory in this N-dimensional space and we use PCA to outline the subspace in which this trajectory lies. To conduct PCA, we diagonalize the equal-time cross-correlation matrix of the firing rates of the \textit{N} units as
\begin{equation}
D_{ij} = <(r_i(t)-<r_i>)(r_j(t)-<r_j>)> 
\end{equation}
where the angle brackets, $<>$, denote time average and $r(t)$ denotes the firing rate activity of the neuron. The eigenvalues of the matrix $D$ (specifically, $\lambda_a/\sum_{a=1}^{N}\lambda_a$, where $\lambda_a$ is the eigenvalue corresponding to principal component $a$) indicate the contribution of different Principal Components (PCs) toward the fluctuations/total variance in the spontaneous activity of the network. 
Fig. S1 shows the impact of varying input amplitude ($I_0$) on the spontaneous chaotic activity of the network. For $I_0 = 0$, the network is completely chaotic as is evident from the highly variable projections of the network activity onto different Principal Components (PCs) as shown in Fig. S1 (a, Left).
Generally, the leading $10-15\%$ (depending upon the value of $g$) of the PCs account for $\sim 95\%$ of the network's chaotic activity \cite{rajan2010inferring}. Visualizing the network activity in a 3D space composed of the dominant principal components (PC1, 2, 3) shows a random and irregular trajectory characteristic of chaos (Fig. S1 (a, Middle). In fact, plotting the trajectories (firing rate $r(t)$ of the neuron as time evolves) of 5 recurrent units in the network (Fig. S1 (a, Right) shows diverging and incoherent activity across 10 different trials, also, representative of intrinsic chaos. In addition, the projections of the network activity onto components with smaller variances fluctuate more rapidly and irregularly (Fig. S1 (a, Left)). This further corroborates the fact that the leading PCs define a network's spontaneous chaotic activity. 

Driving the recurrent network with a sinusoidal input of high amplitude (Fig. S1 (b)) sensitizes the network toward the input, thereby, suppressing the intrinsic chaotic fluctuations. The PC projections of the network activity are relatively periodic. 
A noteworthy observation here is that the trajectories of the recurrent units (Fig. S1 (b, Right) become more stable and consistent across 10 different presentations of the input pattern with increasing amplitude. A readout layer appended to a recurrent network can be easily trained on these stable trajectories for a particular task. Thus, the input amplitude determines the network's encoding trajectories and in turn, its' inference ability. In fact, the chaotic intrinsic activity is completely suppressed for larger inputs. However, this is not preferred as input dominance drastically declines the discriminative ability of a network that can be justified by dimensionality measurements. The effective dimensionality of a reservoir is calculated as $N_{eff} = \sum_{a=1}^{N}(\lambda_a^2)^{-1}$ that provides a measure of the effective number of PCs describing a network's activity for a given input stimulus condition. Fig. S1 (c) illustrates how the effective dimensionality decreases with increasing input amplitude for different $g$ values. It is, hence, critical that input drive be strong enough to influence network activity while not overriding the intrinsic chaotic dynamics to enable the network to operate at the \textit{edge of chaos}. Note, higher $g$ in Fig. S1 (c) yields a larger dimensionality due to richer chaotic activity. 

In our simulations in Fig. S1 (b), the input is shown for $50 ms$ starting at $t =0$. Thus, we observe that the trajectories of the recurrent units are chaotic until the input is turned on. Although the network returns to spontaneous chaotic fluctuations when the input is turned off (at $t = 50 ms$), we observe that the network trajectories are stable and non-chaotic that is in coherence with the previous findings from \cite{bertschinger2004real,rajan2010stimulus}. From the visualization of network activity in the dominant PC space, we see that the input-driven trajectory converges to a uniform shape becoming more circular (along PC1 and PC2 dimensions) with higher input amplitude (Fig. S1 (b)). This informs us that the orbit describing the network activity in the input-driven state consists of a circle in a two-dimensional subspace of the full N-dimensional hyperspace of the neuronal activities. Note, all simulations in supplementary are conducted with similar parameters mentioned in the manuscript, i.e., $ N=800, f= 10Hz, p_c =0.1$.

\section{Mean Field Derivation with Taylor Series}
First, let us solve for ${x_i}^1$ such that we can get an expression for the correlation function, $C(\tau)$ in Eqn. 3 of main manuscript. Noting that, ${x_i}^1$ is driven by Gaussian noise (as indicated by the MFT noise-interaction equation: $\frac{d{x_i}^1}{dt} = -{x_i}^1 +\eta$), we can assume their moments as $<{x_i}^1(t)> = <{x_i}^1(t + \tau)> = 0$, $<{x_i}^1(t) {x_i}^1(t)> = <{x_i}^1(t + \tau) {x_i}^1(t + \tau)> = \Delta(0)$ and $<{x_i}^1(t) {x_i}^1(t + \tau)> =\Delta(\tau)$. $x^1(t)$ (dropping index $i$ as all neuronal variables have similar statistics) can then be written as
 \begin{equation}
x^1(t) = \alpha z_1 + \beta z_3; \hspace{3mm}
x^1(t+\tau) = \alpha z_2 + \gamma z_3
\label{eq6}
\end{equation}
where $z_1, z_2, z_3$ are Gaussian random variables with 0 mean/unit variance and $\alpha = \sqrt{\Delta(0) - |\Delta(\tau)|}, \beta = sgn(\Delta(\tau))\sqrt{|\Delta(\tau)|},\\ \gamma= \sqrt{|\Delta(\tau)|}$. Now, writing $x=x^0 + x^1$, $C$ is computed by integrating over $z_1, z_2, z_3$ as
\begin{eqnarray}
C(\tau) = && 1/N \sum_{i=1}^{N} <<\phi ({x_i}^0(t) 
+ \alpha z_1 + \beta z_3)>_{z_1}\nonumber \\
 && <\phi ({x_i}^0(t+\tau)) + \alpha z_2 
 + \gamma z_3>_{z_2}>_{z_3}
\label{eq7}
\end{eqnarray}
where $<f(z)>_z = \int_{- \infty}^{\infty} dz \frac{f(z) exp(-z^2/2)}{\sqrt{2\pi}} $ for $z = z_1, z_2, z_3$. Now, ${x_i}^0 (t) = A cos(2\pi ft + \zeta)$, where $A = I_0/\sqrt{1+(2\pi ft)^2}$ (solve $\frac{d{x_i}^0}{dt} = -{x_i}^0 +I_0 cos(2 \pi ft + \zeta)$ for ${x_i}^0$) and $\zeta$ incorporates the averaged temporal phase relationship between the individual neurons and the input induced by input subspace alignment. Replacing the value of ${x_i}^0$ in Eqn. \ref{eq7}, we get 
\begin{eqnarray}
C(\tau) = && 1/N \sum_{i=1}^{N} <<<\phi (Acos(\zeta)+ \alpha z_1 + \beta z_3)>_{z_1}) \nonumber \\
&& <\phi(Acos(\omega \tau + \zeta) + \alpha z_2 + \gamma z_3>_{z_2}>_{z_3})>_\zeta
\label{eq8}
\end{eqnarray}
The above correlation function also satisfies Eqn. 4 of main manuscript. Note, $\omega = 2\pi f$ in Eqn. \ref{eq8}. 

Now we solve Eqn. \ref{eq8} using Taylor series approximation for $tanh(gx) = gx-1/3g^3x^3 +2/15g^5x^5$. We have
\begin{eqnarray}
\phi(Acos(\zeta) + \alpha z_1 +\beta z_3)= &&g(Acos(\zeta) + \alpha z_1 +\beta z_3)-1/3g^3(\alpha z_1+\beta z_3)^3+2/15g^5(\alpha z_1 + \beta z_3)^5 \nonumber \\
=&& g(Acos(\zeta) + \alpha z_1 + \beta z_3) -1/3g^3(\alpha^3 {z_1}^3 + 3 \alpha^2 {z_1}^2 \beta z_3 + 3 \alpha z_1\beta^2 {z_3}^2 + \beta^3 {z_3}^3) \nonumber\\
&& +2/15 g^5(\alpha^5 {z_1}^5 + 5 \alpha^4 {z_1}^4 \beta z_3 + 10 \alpha^3 {z_1}^3 \beta^2 {z_3}^2 +10\alpha^2 {z_1}^2 \beta^3 {z_3}^3 \nonumber\\
&&+5 \alpha z_1 \beta^4{z_3}^4+\beta^5{z_3}^5
\label{eq9}
\end{eqnarray}
Now using $<{z_1}^2>_{z_1} =1, <{z_1}^4>_{z_1} =3, <{z_1}^6>_{z_1} =15$ and noting that averages over odd powers of $z_1$ are zero, we get
 \begin{eqnarray}
<\phi(Acos(\zeta) + \alpha z_1 +\beta z_3)>_{z_1}= &&g(Acos(\zeta) +\beta z_3)-1/3g^3(3 \alpha^2 \gamma z_3 +\gamma^3 {z_3}^3)\nonumber \\
&&+ 2/15g^5(15\alpha^4 \beta z_3 + 10 \alpha^2\beta^3 {z_3}^3 +\beta^5 {z_3}^5) 
\label{eq10}
\end{eqnarray}
Similarly,
 \begin{eqnarray}
<\phi(Acos(\zeta +\omega \tau) + \alpha z_2 +\gamma z_3)>_{z_2}= &&g(Acos(\zeta +\omega \tau) +\gamma z_3)-1/3g^3(3 \alpha^2 \gamma z_3 +\gamma^3 {z_3}^3)\nonumber \\
&&+ 2/15g^5(15\alpha^4 \gamma z_3 + 10 \alpha^2\gamma^3 {z_3}^3 +\gamma^5 {z_3}^5) 
\label{eq11}
\end{eqnarray}

Multiplying Eqn. \ref{eq10} with Eqn. \ref{eq11} upto the sixth order, we can simplify Eqn. \ref{eq8} as
 \begin{eqnarray}
C =&&<<g^2 A^2 cos(\zeta) cos(\zeta +\omega \tau) + g^2 \beta \gamma {z_2}^2 -2 g^4\alpha^2 \beta \gamma {z_3}^2 -1/3 g^4(\beta^3 \gamma +\beta \gamma^3){z_3}^2\nonumber \\
&&+ 4g^6 \alpha^4 \beta \gamma {z_3}^2 + 4/3 g^6 \alpha^2(\beta^3\gamma + \beta \gamma^3){z_3}^4 +2/15 g^6(\beta^5 \gamma +\beta \gamma^5){z_3}^6 \nonumber \\
&& +g^6\alpha^4 \beta \gamma {z_3}^2+1/3g^6\alpha^2(\beta^3\gamma+\beta\gamma^3){z_3}^4+1/9g^6\beta^3\gamma^3{z_3}^6>_{z_3}>_\zeta 
\label{eq12}
\end{eqnarray}
Now averaging Eqn. \ref{eq12} over $\zeta$ will still retain the $\zeta$ term unlike general MFT calculations where $\zeta$ gets washed out since there is no fixed relationship between the input and recurrent activity. Here, we use yet another approximation i.e. $Cos(A). Cos(B) \simeq Cos(A+B)$ and finally get $C$ as
\begin{eqnarray}
C =&&1/2g^2 A^2 cos(2\zeta +\omega \tau) +g^2 \beta \gamma - 2 g^4\alpha^2 \beta \gamma - g^4 (\beta^3\gamma + \beta \gamma^3) +5g^6\alpha^4\beta\gamma \nonumber\\
&&+5g^6\alpha^2(\beta^3\gamma + \beta \gamma^3) +2g^6(\beta^5 \gamma +\beta\gamma^5)+5/3 g^6 \beta^3 \gamma^3
\label{eq13}
\end{eqnarray}
Using the definition of $\alpha, \beta, \gamma$ in Eqn. \ref{eq6} and denoting $m=\Delta(0), l=\Delta(\tau)$, we get
\begin{eqnarray}
\alpha^2=m-|l|;\hspace{1mm} \beta\gamma=l;\hspace{1mm} \beta^3\gamma=\beta\gamma^2=|l|l; \hspace{1mm} \beta^5\gamma=\beta\gamma^5=\beta^3\gamma^3=l^3
\label{eq14_1}
\end{eqnarray}
Substituting $\alpha, \beta, \gamma$ values from Eqn. \ref{eq14_1} in Eqn. \ref{eq13}, we get
\begin{eqnarray}
C =&&1/2g^2 A^2 cos(2\zeta +\omega \tau) +g^2 l-2g^4(m-|l|)l-2g^4|l|l+5g^6(m-|l|^2l +10g^6(m-|l|)|l|l+17/3g^6l^3 \nonumber\\
=&& 1/2g^2A^2cos(\omega \tau +2\zeta) + l(g^2-2g^4m) + 2/3g^6l^3
\label{eq14}
\end{eqnarray}
Now, $l$ satisfies $\frac{d^2l}{dt^2}=l-C$. By putting the value of $C$ from Eqn. \ref{eq14} we get
\begin{eqnarray}
\frac{d^2l}{dt^2}=&& 1/2g^2A^2cos(\omega \tau +2\zeta) + (1-g^2-2g^4m)l + 2/3g^6l^3
\label{eq15_1}
\end{eqnarray}
For deriving Eqn.5 from the main manuscript, we use $l=k\delta$ and define $1/2g^2A^2cos(\omega \tau +2\zeta) = G\delta^3$. Then, Eqn. \ref{eq15_1} becomes
\begin{eqnarray}
\delta^3\frac{d^2k}{dt^2}=&& -\delta^3G+\delta(-2\delta+2(1+4\delta)m)k-2/3\delta^3k^3
\label{eq15}
\end{eqnarray}
Now, we introduce the parameter $n$, where $m = \delta +(n/2-4)\delta^2$ to further simplify Eqn. \ref{eq15} as
\begin{eqnarray}
\frac{d^2k}{dt^2} = -G + n k -2/3k^3
\label{eq16}
\end{eqnarray}
Eqn. \ref{eq16} is the force equation (Eqn. 5 in the manuscript) from which we derived the potential energy. Note, the MFT approximate equations are derived with respect to a single input $G$ driving the entire network. Multiple inputs ($G_{in}$ where $in=1,2...$) and corresponding alignment of the inputs along different projections (as in Fig. 1 of main manuscript) will result in a potential well that can be roughly interpreted as a linear combination of $V_{in}$ observed for each input, $G_{in}$. The linear combination will follow a similar evolution profile as shown in Fig. 2 of main manuscript. Thus, the change in intra-trajectory distance for varying alignment of the two inputs in the chaotic subspace, shown in Fig. 1 (f) of main manuscript, is justified from the given analysis.

\section{Handwriting Generation}
To further elucidate the effectiveness of input alignment for complex pattern generation, we trained a recurrent network to generate the handwritten words ``chaos'' and ``neuron'' in response to two different inputs \cite{laje2013robust, S5}. After obtaining the principal angle of the chaotic spontaneous activity, we aligned the input $I_1$ corresponding to ``chaos'' along $\angle PC1.PC2$ using optimal input phase $\Theta$. Then, we monitored the output activity for different orientation (i.e. $90^\circ, 180^\circ$) of input $I_2$, corresponding to ``neuron'', with respect to $I_1$ in the chaotic subspace. The two output units (representing the $x$ and $y$ axes) were trained using RLS to trace the original target locations ($x(t), y(t)$) of the handwritten patterns at each time instant. Fig. S2 (a) shows the handwritten patterns generated by the network across 10 test trials for the scenario when inputs are aligned at $90^\circ$ in the chaotic subspace. We observe similar robust patterns generated for $180^\circ$ phase as well. The notable feature of input alignment is that the chaotic trajectories become locally stable channels and function as dynamic attractor states. This can be visualized from the stable and synchronized trajectories observed for different neural units in Fig. 1 (b, Right) of main manuscript (and in Fig. S1). However, external perturbation can induce more chaos in the reservoir that will overwhelm the stable patterns of activity. 

To test the susceptibility of the dynamic attractor states formed with input structuring to external perturbation, we introduced random Gaussian noise onto a trained model along with the standard intrinsic chaos-aligned inputs during testing. The injection of noise alters the net external current received by the neuronal units ($I = \Sigma_i[W_{{Input}_i} I + N_0 rand(i)]$, where $N_0$ is the noise amplitude, $i$ denotes a neural unit in the reservoir and $rand$ is the random Gaussian distribution). Fig. S2 (b) shows the mean squared error (calculated as the average Euclidean distance between the target ($x, y$) and the actual output produced at different time instants, averaged across 20 test trials) of the network for varying levels of noise. As $N_0$ increases, we observe a steady increase in the error value implying degradation in the prediction capability of the network. However, for moderate noise (with $N_0 <0.01$), the network exhibits high robustness with negligible degradation in prediction capability for both the words. Interestingly, for $90^\circ$ phase difference, ``neuron'' is more stable than ``chaos'' with increased reproducibility across different trials even with more noise ($N_0 =0.2$). In contrast, for $180^\circ$ phase, ``chaos'' is less sensitive to noise (Fig. S2 (b)). On the other hand, for $45^\circ$ phase alignment between $I_1, I_2$ in the chaotic subspace, we observe that the network is sensitive even toward slight perturbation ($N_0 = 0.001$). This implies that the attractor states formed, in this case, are very unstable. This further corroborates the fact that the extent of noise suppression and hence the attractor state stability varies based upon the input alignment. 

\begin{figure}
\centering
\includegraphics[width=0.8\linewidth]{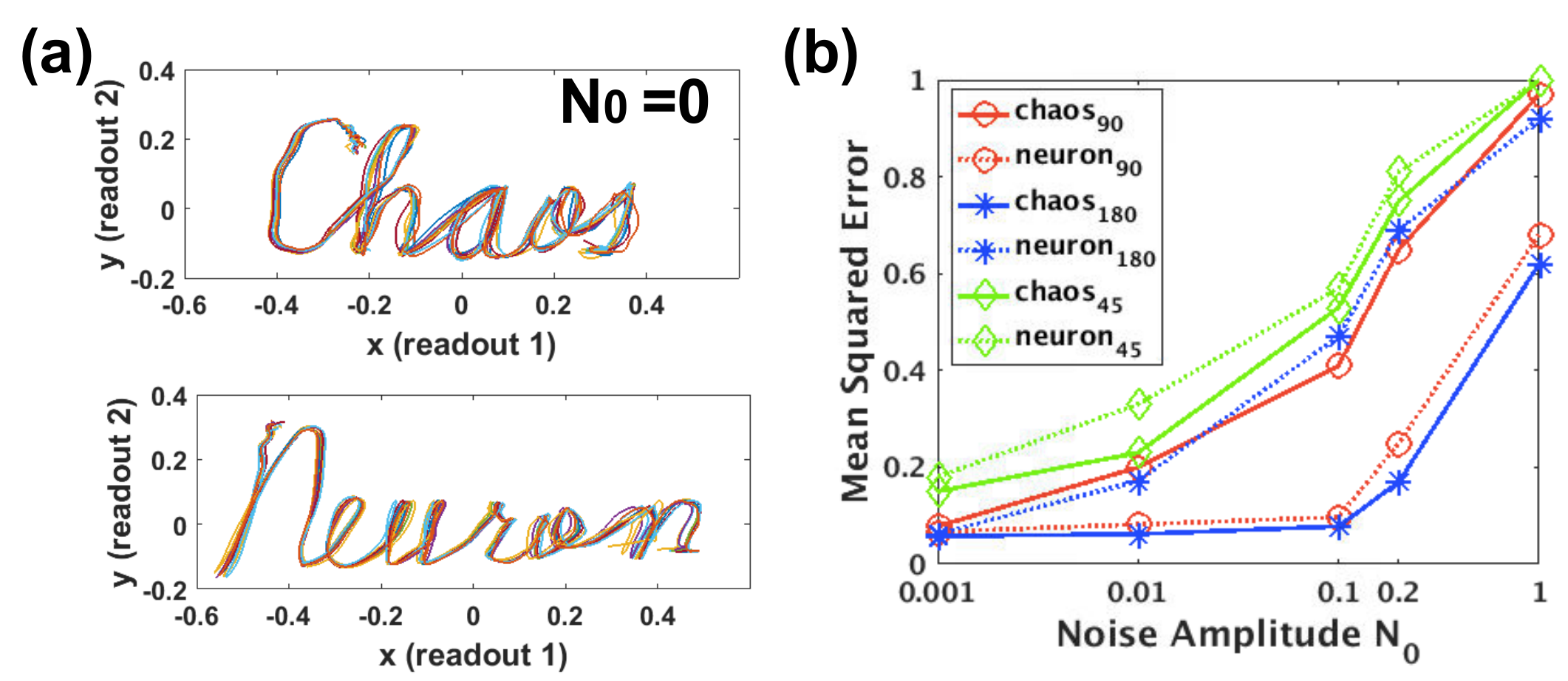}
\caption*{\textbf{Figure S2} (a) Handwriting patterns generated across 10 test trials in response to $I_1$ [Top] and $I_2$ [Bottom] in absence of external noise (b) Variation of performance (measured as mean squared error) with different noise amplitude shown for each output pattern in different input alignment scenarios. $Chaos_{90,180,45}$ represents the error value obtained during ``chaos'' generation when $I_1, I_2$ have $90^\circ, 180^\circ, 45^\circ$ phase difference in the chaotic subspace, respectively.}
\vspace{-4mm}
\end{figure}

\begin{figure}
\centering
\includegraphics[width = 0.9\linewidth]{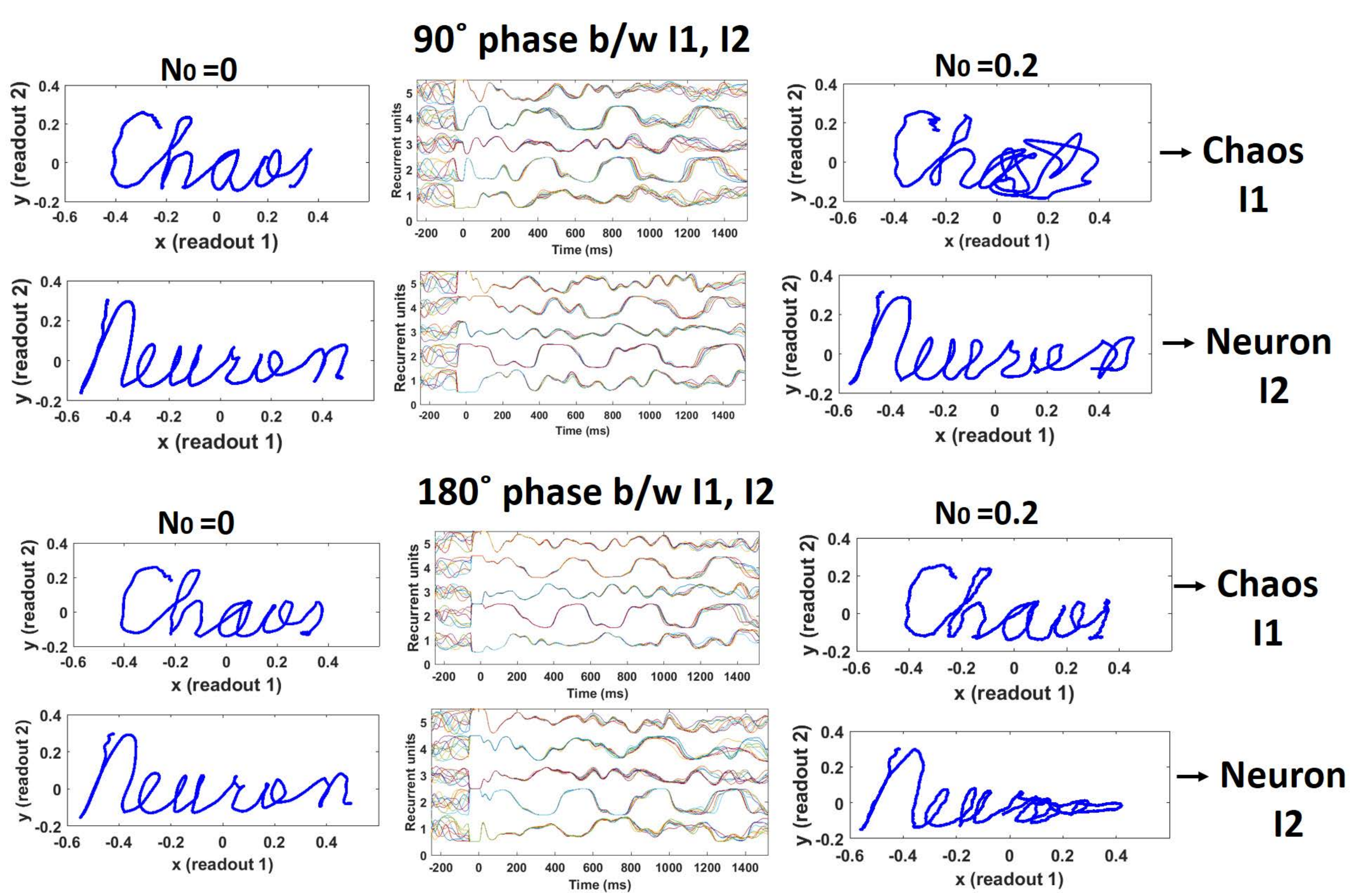}
\caption*{\textbf{Figure S3} Generation of handwriting patterns corresponding to two brief sinusoidal inputs $I_1, I_2$ for different phase difference between the inputs in the chaotic subspace. [Left column] shows the patterns generated in absence of external perturbation in one test trial, [Middle Column] shows the trajectories of 5 recurrent units in the reservoir across 10 test trials corresponding to each output pattern, [Right Column] shows the patterns generated in presence external perturbation for one test trial.}
\end{figure}
Fig. S3 shows the handwritten pattern generated in one test trial for different phase alignment between $I_1, I_2$, when $I_1$ is aligned along the principal angle defining the spontaneous chaotic activity of the network. It is noteworthy to mention that the neural trajectories of the recurrent units corresponding to both cases are stable. In fact, we observe in the $90^\circ$ case, the trajectories of neurons responding to $I_1$ that corresponds to output ``chaos'' become slightly divergent and incoherent beyond $1000ms$. In contrast, the trajectories of units responding to the word ``neuron'' are more synergized and coherent throughout the $1500 ms$ time period of simulation. This indicates that the network activity for ``neuron'' converges to a more stable attractor state than ``chaos''. As a result, we see that the network is more robust while reproducing ``neuron'' even in presence of external perturbation ($N_0$, noise amplitude is 0.2). In the $180^\circ$ phase difference case, we see exactly opposite stability phenomena with ``chaos'' converging to more stable attractor.

\end{document}